\documentclass[11pt,twoside]{article}
\usepackage{cal10}
\usepackage{color}
\usepackage{graphicx}

\contact{Steven Osterman}
\email{steven.osterman@colorado.edu}

\begin{document}

%
%
%

\title{Observing with HST below 1150\AA: Extending the Cosmic Origins 
	Spectrograph Coverage to 900\AA}
\titlemark{Observing with HST below 1150\AA}

%
%
%

\author{Steve Osterman, Steven V. Penton, 
	Kevin France, St\'{e}phane B\'{e}land}
\affil{Center for Astrophysics and Space Astronomy, Astrophysics Research Lab, 
		University of Colorado, Boulder, CO, 80309}
\author{Stephan McCandliss,}
\affil{Department of Physics and Astronomy, Johns Hopkins University, 
		Baltimore, MD 21218}

\author{Jason McPhate}
\affil{University of California, Berkeley Space Sciences Lab, Berkeley, CA 94720}

\author{Derck Massa}
\affil{Space Telescope Science Institiute, Baltimore, MD 21218}


%
%
%

\paindex{Osterman, S.}
\aindex{Penton, S.}
\aindex{France, K.}
\aindex{B\'{e}land, S.}
\aindex{McCandliss, S.}
\aindex{McPhate, J.}
\aindex{Massa, D.}

%
%

\authormark{OSTERMAN et al.}



\begin{abstract}
The far-ultraviolet (FUV) channel of the Cosmic Origins Spectrograph (COS) 
is designed to operate between 1130\AA~and 1850\AA, limited at shorter wavelengths 
by the reflectivity of the MgF$_{2}$ protected aluminum reflective surfaces on the Optical Telescope 
Assembly and on the COS FUV diffraction gratings. However, because the detector 
for the FUV channel is windowless, it was recognized early in the design phase 
that there was the possibility that COS would retain some sensitivity at shorter 
wavelengths due to the first surface reflection from the MgF$_{2}$ coated optics. Preflight 
testing of the flight spare G140L grating revealed $\sim$ 5\% efficiency at 1066\AA, 
and early on-orbit observations verified that the COS G140L/1230 mode was sensitive 
down to at least the Lyman limit with 10-20 cm$^{2}$ effective area between 912\AA~and 1070\AA, 
and rising rapidly to over 1000 cm$^{2}$ beyond 1150\AA.   Following this initial work we 
explored the possibility of using the G130M grating out of band to provide coverage 
down to 900\AA. We present calibration results and ray trace simulations for these 
observing modes and explore additional configurations that have the potential to 
increase spectroscopic resolution, signal to noise, and observational efficiency 
below 1130\AA.
\end{abstract}
\vspace{-0.2cm}
\section{Introduction}
The Cosmic Origins Spectrograph (COS), installed in the Hubble Space Telescope in May, 2009, 
was intended to provide 
high sensitivity, moderate to low resolution spectroscopy between 1130\AA~and 3200\AA~(Green 2003). 
In addition to meeting this goal, COS has demonstrated sensitivity down to wavelengths 
approaching 900\AA~(McCandliss 2010), providing coverage in the Far Ultraviolet Spectroscopic 
Explorer (FUSE) band at sensitivities comparable to individual FUSE channels. The nominal G140L/1230 
mode provides coverage from the detector cutoff at 1850\AA~down to  $<$910\AA, and 
two new modes, G130M/1096 and /1055, provide higher sensitivities and potentially much higher 
signal to noise at these wavelengths.

\begin{figure}
\plotone{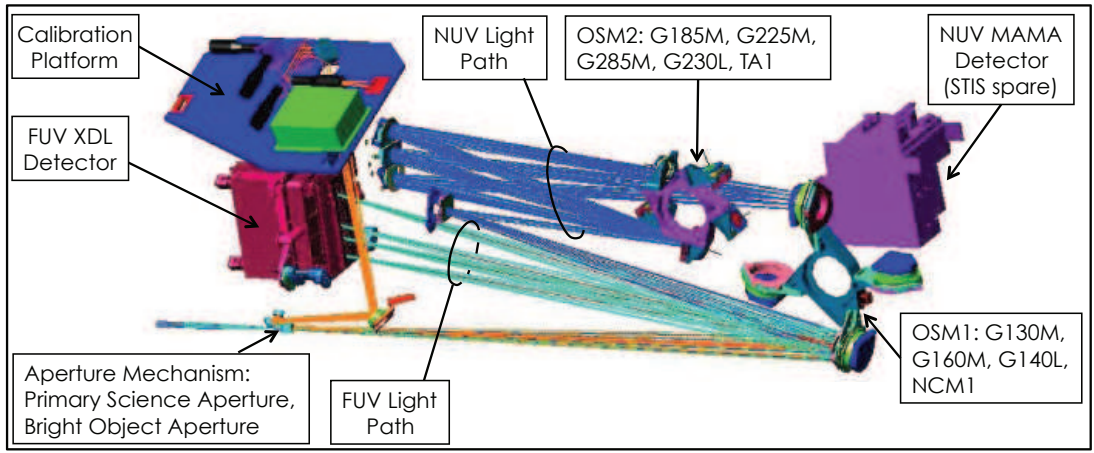}
\caption{COS Light Path.  The FUV channels are distinguished by 
having only one reflection and a windowless, two segment detector.  
The holographically-ruled aspheric 
FUV diffraction gratings provide dispersion, reimaging, and astigmatism 
 and aberration correction in a single reflection.  The two segment
detector allows for one segment to be disabled to prevent overlight if asymmetric
illumination is anticipated. 
(Froning 2009)}
\label{fig:lightpath}
\end{figure}

\begin{figure}
\plotone{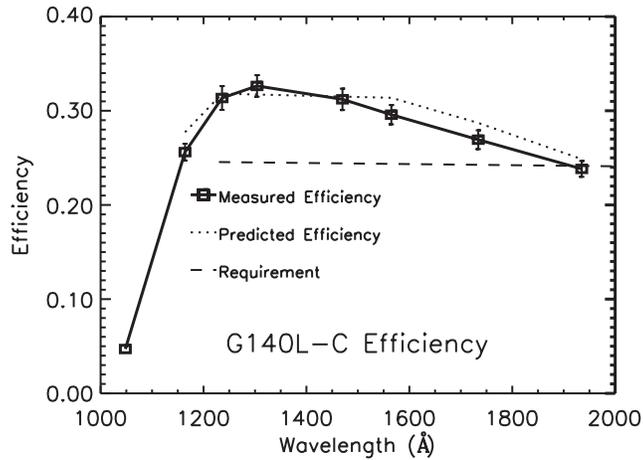}
\caption{G140L-C (flight spare) laboratory test results showing $\sim$5\% total efficiency 
(groove efficiency$\times$reflectivity) at 1066\AA. }
\label{fig:g140leff}
\end{figure}

The COS light path is shown in figure \ref{fig:lightpath}. 
Light from the Optical Telescope Assembly (OTA) enters COS through an oversized 
aperture admitting the entire aberrated wavefront from a point-like source. The 
aperture is windowless so that any short wavelength ($>1130$\AA) light remaining after 
the two reflections in the OTA will travel unobstructed to the FUV grating. The 
gratings perform diffraction, aberration correction and focus in a single reflection 
in order to minimize reflections, maximizing short wavelength sensitivity. 
Light then travels to the windowless FUV detector. Laboratory testing indicated that 
the detector retains relatively high ($>$30\%) quantum efficiency down to at least 800\AA. 
While the COS diffraction gratings and the OTA mirrors are coated with MgF$_{2}$ protected aluminum 
(typically used for wavelengths longer than 1150\AA), 
these optics were expected to retain some first-surface reflectivity below the MgF$_{2}$ transmission 
cutoff at $\sim$1150\AA. This was verified for the G140L-C (flight spare) grating in laboratory 
testing (fig.\ref{fig:g140leff}) (Osterman 2002), and for the OTA during COS on-orbit testing.

\vspace{-0.4cm}
\section{G140L Modes}
\subsection{G140L/1230}
\begin{figure}
\plotone{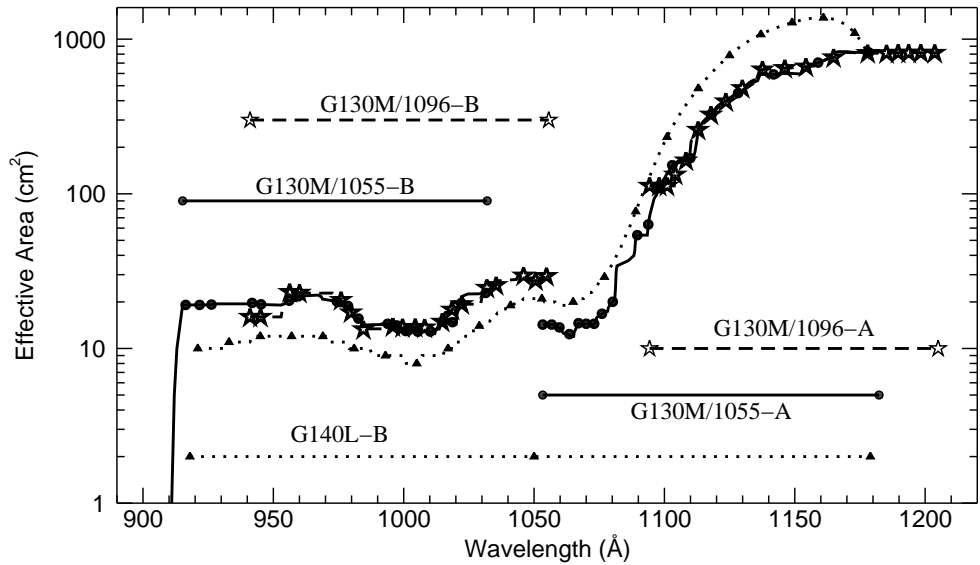}
\caption{Effective area of G140L and G130M modes at wavelengths below 
1200\AA. The G140L/1230-B segment (dotted line + triangles) provides coverage 
from $>$920\AA -1170\AA\ with sensitivity increasing rapidly above 1075\AA. Instrument 
configurations using the G130M for sub-1150\AA\ observations (solid line + circles, 
dashed line + stars) provide narrower coverage per detector segment. 
By placing the long wavelength edge of the G130M B segment below 1075\AA\
we avoid the large variation in sensitivity across a single detector segment evidenced in the G140L/1230 
mode. This permits observations of brighter objects, enabling higher 
signal to noise observations at shorter wavelengths.  FUSE effective 
area ranged from $\sim7-25$ cm$^{2}$ per channel (Sahnow 2000). }
\label{fig:aeff}
\end{figure}

The FUV detector on COS is composed of two independently commandable $10\times85$mm 
segments, referred to as segments A and B (B is the shorter wavelength segment in all modes). 
Although the B segment is typically maintained a reduced voltage (HV-Low) 
for G140L/1230 
setting, effectively disabling the snort wavelenght half of the detector, segment B was
was intentionally left at the operating voltage (HV-Nom) during the first calibration tests. 
It was immediately 
obvious that the instrument and the OTA retained significant sensitivity down to 
approximately 900\AA. Measured sensitivity is shown in fig. \ref{fig:aeff}, dotted line + triangles).  Spectroscopic 
resolution ($\lambda/\Delta\lambda$) is expected to drop from $\sim$ 2300 at 1250\AA\ 
to no better than 2000 at 900\AA\ (McCandliss 2010). 

\subsection{G140L/800}
The success of the G140L/1230-B observations suggested the possibility of 
shifting the wavelength scale to the blue so that the central wavelength 
(falling on the detector gap) is $\sim$ 800\AA. This would place 
the entire G140L pass band on a single detector segment and eliminate the 
need for multiple observations to obtain full coverage. This requires 
a relatively large focus mechanism move, placing the focus mechanism (discussed below) outside of the 
nominal focus range for the G130M and G160M gratings, but it could be argued 
that this does not represent an unacceptable risk given the increased observational 
efficiency that this mode could provide.

\subsection{Red Leak}
One difficulty encountered in the G140L/1230-B mode (and anticipated for the 
G140L/800 mode) is the greater than two order of magnitude increase in the effective area 
with wavelength across the B segment of the detector, rising 
from $\sim$ 10 cm$^{2}$ at the shortest wavelengths 
to over 1300 cm$^{2}$ at the long wavelength edge. As a result of this sensitivity variation only relatively 
dim targets can be observed without triggering the bright object protection 
despite the low sensitivity at the blue end of the 
band pass, significantly reducing the obtainable signal to noise.  While the G140L 
grating could be repositioned so as to place the higher effective area portions of the spectrum
on the A (disabled) detector segment to obtained more uniform sensitivity, this would bring
zero-order light onto the B segment (as is the case with the G140L/1105 setting), 
undoing any attempts to flatten the countrate on the B segment.

\section{G130M Modes}
The instrument development team had not originally considered any short wavelength 
configurations for the G130M grating because we anticipated significantly degraded 
resolution as we moved farther and farther out of band and because we had no model or test 
data for the grating efficiency at wavelengths far from blaze. However, in light of 
the better than expected performance in the G140L/1230-B segment, we began exploring
the possibility driving the short wavelength cutoff of the G130M spectrum 
down to 900\AA.  This had the potential to mitigate the 
unwieldy variation in sensitivity exhibited in the G140L/1230 mode and expected 
with the G140L/800 mode. 

\subsection{G130M/1055 and /1096}
The COS FUV gratings are located $\sim$178mm from the center of rotation of the 
grating select mechanism (fig. \ref{fig:lightpath}).  In addition to providing grating selection, 
this mechanism permits small adjustments to the band for each grating. This flexibility 
ensures that the full spectrum can be covered despite the gap 
introduced by the physical separation of the A and B detector segments. 
The grating select mechanism is in turn mounted on a linear
translation mechanism to take out motion of the grating along the optical axis
 and to approximately accommodate changes to the location of the focal surface 
 introduced by grating rotation. Larger than nominal moves 
are possible, but require substantial granting mechanism translations 
to correct for the focus offset. Large wavelength offsets may require a greater 
focus adjustment than the mechanism can provide,  
and even then would not fully recover 
the nominal instrument performance since the focal plane will no longer be 
coincident with the detector face. Also, outside of the nominal band the cross 
dispersion height is compromised.

Nevertheless, if reduced resolution is acceptable the gratings can be configured to 
support observations significantly outside of the design wavelength range. 
We proposed observing with the G130M grating rotated so that 900\AA\  light would fall 
on the short wavelength edge of B detector segment and with a second position offset from the first 
by +41\AA\  to ensure
full coverage across the detector gap and overlap with the G130M/1291 mode. 
This configuration requires 
rotating the grating mechanism approximately $2.8^{\circ}$;  the corresponding focus mechanism
position that returns the best possible resolution is  roughly a factor of two beyond the 
mechanism hard stops. Given the shallow slope of line width versus focus mechanism position, 
we chose to limit the focus offset to within the range used by existing modes.  This represents a small
reduction in resolution, but ensures that in the event of a focus mechanism failure during an 
observation in one of these new configurations
 performance would not be as severely degraded in the in the nominal G130M modes as it would with
 the mechanism at the extreme range of motion. 
\begin{figure}
\epsscale{1.05}
\plotone{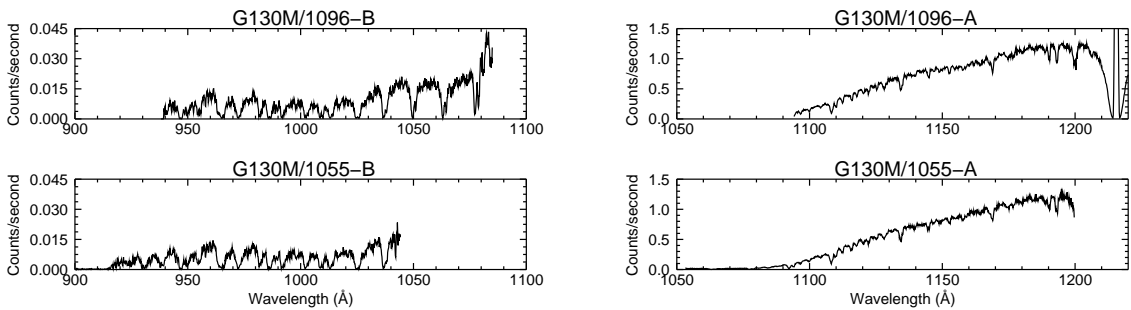}
\caption{G130M/1055 and /1096 observations of GD561 obtained by COS.}
\label{fig:gd561}
\end{figure}

\begin{figure}
\plottwo{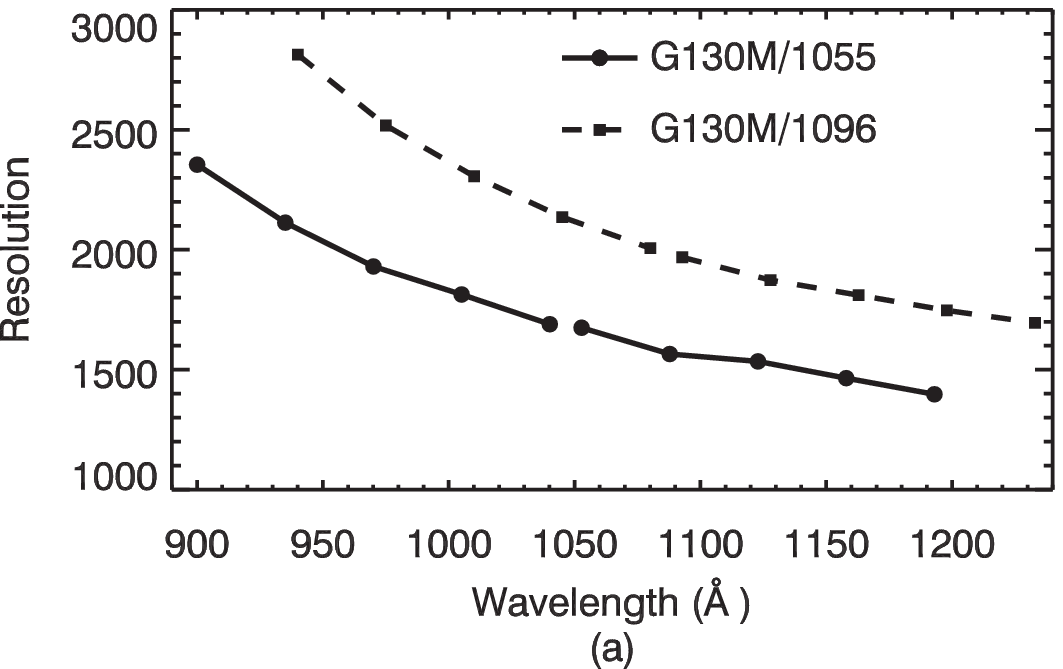}{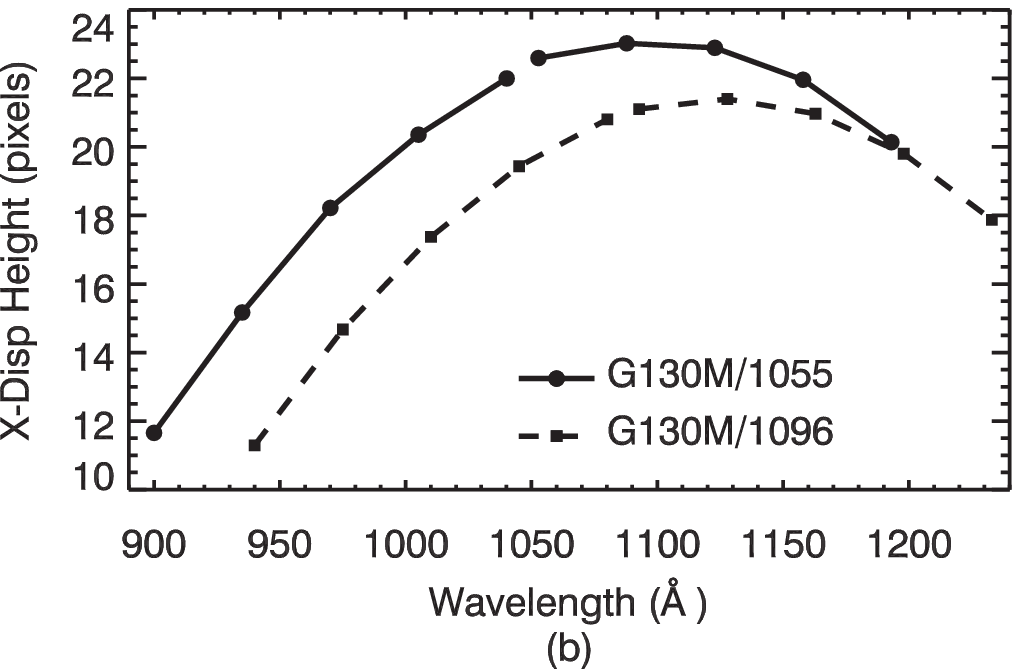}
\caption{Ray trace results for G130M/1055 and /1096 show (a) resolution decreasing at longer 
wavelengths and (b) cross 
dispersion width peaking near the central wavelength. Pixels are approximately 
6$\mu$m wide (dispersion direction) by 25$\mu$m tall.  OTA mid frequency wavefront 
error is not included in the geometrical ray trace model so that actual resolution is expected
to be somewhat lower than model predictions.}
\label{fig:g130mraytrace}
\end{figure}

The two new modes, G130M/1055 and /1096, provide continuous spectral 
coverage from 900\AA\ to the short wavelength edge of the existing G130M modes. 
Observations of GD561 were carried out over the third quarter of 2010 
(fig. \ref{fig:gd561}) and the effective 
area was determined to be higher than for the G140L/1230-B mode (fig. \ref{fig:aeff}). 
Spectral resolution in these modes is under evaluation and appears to be consistent with
 ray trace predictions,
dropping from approximately 2500 at the shortest wavelengths to 1400 at longer wavelengths 
(fig.  \ref{fig:g130mraytrace} and table \ref{table:summary}).  While the large sensitivity variation 
is still apparent, with the G130M/1055 and /1096 modes the A segment of the detector
can be disabled permitting observation of brighter objects to obtain higher S/N observations.

\subsection{G130M-1215}

In addition to the G130M/1055 and /1096 modes, we have performed ray trace 
modeling of a third new G130M mode, extending short wavelength coverage down to 
1065\AA. This mode does not require the large grating rotation and corresponding 
large refocusing move demanded by the G130M/1055 and /1096 configurations 
as the short wavelength edge is now only 55\AA\ beyond the blue edge of G130M-1291. 
This off-nominal geometry compromises the best possible resolution and cross 
dispersion height, but to a much lesser degree than for the more extreme cases already 
explored, maintaining better than 16,000 resolution across the band.  
This mode has the added advantage of placing geocoronal Ly-$\alpha$ on the detector gap.

\vspace{-0.2cm}
\section{Conclusion}
The Hubble Space Telescope has provided spectacular imaging and spectroscopy 
longwards of 1150\AA\ for over 20 years. The Cosmic Origins Spectrograph, while 
originally intended to provide spectroscopic coverage from 1130\AA\ to 3200\AA, has 
now extended the usable wavelength range of HST to the Lyman limit, providing 
spectroscopic access to wavelengths unobservable since the end of the FUSE mission 
in 2007.  The capabilities of these new and proposed modes are summarized in 
table \ref{table:summary}. 

\begin{table}
\caption{New and Proposed COS Observing Modes \label{table:summary}}
\centering
\begin{tabular}{|c|c|c|c|c|}
\hline
COS                  & Wavelength       & Effective         & Modeled               & Background \\
Mode               & Range                 &  Area                 & Resolution    & (cts/resl/ksec) \\
\hline
\underline{G140L/1230} & $<$920-1160\AA & $\sim$8-10 cm$^{2}$ & $\sim$2100 & 0.3 \\
                         & 1230-1850\AA & (at 1000\AA)                      &  (1000\AA)           &         \\
\hline
{\it G140L/800}   & $<$920-1850\AA & $\sim$8-10 cm$^{2}$ & $\sim$2100 & 0.2 \\
                                   & (a segment) & (at 1000\AA)                      &  (1000\AA)             &         \\
\hline
\underline{G130M/1096}  & 940-1081\AA & $\sim$15-25 cm$^{2}$ & $\sim$2300 & 1.3 \\
                             & 1096-1238\AA & (b segment)                     &  (1000\AA)            &         \\
\hline
\underline{G130M/1055}  & 900-1041\AA & $\sim$15-25 cm$^{2}$ & $\sim$1800 & 1.8 \\
                             & 1055-1196\AA & (b segment)                     &  (1000\AA)            &         \\
\hline
{\it G130M/1215}  & 1065-1205\AA & $\sim$30-2000 cm$^{2}$ &  $\sim$16,000 & 0.1 \\
                             & 1220-1360\AA & (b segment)                      &  (1100\AA)                          &         \\
\hline
\multicolumn{5}{|l|}{Underlined modes are available or will be made available to observers in}\\
\multicolumn{5}{|l|}{cycle 19. Modes in italic have not been tested and performance projections }\\
\multicolumn{5}{|l|}{are based on modeling and on similar modes.}\\
\hline
\end{tabular}
\end{table}

By expanding coverage to these shorter wavelengths, we make possible a range of 
investigations not previously accessible to HST, including studies of the Lyman 
continuum escape fraction from low redshift galaxies, of the Gunn Peterson effect at redshifts 
between 2 and 2.8 along multiple lines of sight, and observations of the O VI $\lambda\lambda$ 
1032, 1038 doublet. 
The higher resolution of the G130M/1215 mode could support 
observation of atomic and molecular diagnostics 
that could be used to study winds and atmospheres of massive stars, as well as the 
bulk of the mass in the translucent ISM out of which those stars form.


\vspace{-0.2cm}
\begin{references}
\reference Froning, C.~S., \& Green, J.~C.\ 2009, \apss, 320, 181
\reference Green, J.~C., Wilkinson,  E., \& Morse, J.~A.\ 2003, Proc SPIE, 4854, 72 
\reference McCandliss, S.~R.,  France, K., Osterman, S., Green, J.~C., McPhate, J.~B.,
			\& Wilkinson, E.\ 2010, \apjl, 709, L183
\reference Osterman, S.~N.,  Wilkinson, E., Green, J.~C., \& Redman, K.~W.\ 2002,
			Proc SPIE, 4485, 361   
\reference Sahnow, D.~J., et al.\  2000, Proc SPIE, 4013, 334   
\reference Sahnow, D.~J., et al.\ 2010, Proc SPIE, 7731
\end{references}
\end{document}